          [2004/06/22 1.2 (KOF); 2001/04/25 1.1 (PWD)]

\documentclass[a4paper,twocolumn]{Gaia2004} 
\usepackage{times}      
\usepackage{epsfig}     
\usepackage{natbib}     
\title{WEIGHING STELLAR--MASS BLACK HOLES WITH GAIA}

\author{B. Fuchs , U. Bastian}
\affil{Astronomisches Rechen--Institut, 69120 Heidelberg, Germany}

\bibpunct{(}{)}{;}{a}{}{,}  

\begin{document}


\maketitle

\begin{abstract}

 Stellar-mass black holes have been detected by radial-velocity observations in
 star/black hole binaries. These allow only the determination of the mass 
 function. Tracking the astrometric orbits of the visible components of the
 star/black hole binaries would allow the full determination of the black hole 
 masses, which would be of great astrophysical interest. We investigate the 
 possibilities to do this with Gaia. A very promising object seems to be 
 Cyg X-1. The donor star is an O9.7Iab supergiant having a mass of 25 $M_\odot$,
 and the mass of the black hole is estimated to be at least 13 $M_\odot$ 
 masses. The period of 5.6 days implies a semi-major axis of 1.01 $\mu$pc of 
 the binary orbit. At the distance of Cyg X-1 (2.5kpc) this translates into a 
 semi-major axis of about 28 $\mu$arcsec for the visible component, 
 which can be detected by Gaia. We discuss further candidates. 

\end{abstract}

\section{Introduction}

Candidates for stellar--mass black holes have preferentially been found as X-ray
sources in binary systems by radial velocity measurements of the optically
visible companion star. Such measurements allow, however, only the
determination of the mass functions
\begin{equation}
\label{eq1}
f(M) = {Pa^3_{RV} \over 2 \pi G} = {M^3_{BH}~sin~i^3 \over (M_{BH} + M_*)^2} \,,
\end{equation}
where $P$ denotes the orbital period, $a_{RV}$ is the semi amplitude of the
companion star's radial velocity curve and $G$ is the constant of gravity.
$M_{BH}$ and $M_*$ are the masses of the black hole and the companion star,
respectively. $i$ denotes the inclination of the orbital plane to the line of
sight in the sense that $i = 90^\circ$, if the orbit is seen edge-on. The
inclination angle is not known from the radial velocity curve. If the
latter is combined with the light curve and a measurement of the projected
rotational velocity broadening of the spectral lines of the companion star, 
$i$~can sometimes be estimated indirectly (Orosz 2003). A direct,
much more accurate determination of the mass of the black hole is possible, if
the orbit of the companion star is observed also astrometrically. A precise
measurement of the black hole mass would help to constrain the radius of the
last stable orbit around the black hole. If the orbital period of
material on such an orbit can be identified by intensity fluctuations of the
system, this would be a physical experiment of great significance, because it
would prove the existence of event horizons and allow the measurement of the
spin parameter of the black hole.

\section{The candidate Cyg~X-1}

We have examined the objects given in the compilations of Orosz (2003) and
Tanaka \& Lewin (1995) and tested whether the 
orbits of the companion stars can be observed with
Gaia. We have found five objects for which the semi-major axis of the
astrometric orbits exceed 10~micro-arcseconds,
although four of them seem to be too faint to be actually measured by Gaia.
Orbits of these four
could in principle be detected by the SIM space interferometer, but
would require a very significant amount of observing time.
The parameters
of the systems as well as distances and mass
estimates for the black holes have been adopted from Orosz (2003), 
Tanaka \& Lewin (1995) and references therein and
are given in Table 1 for the five objects. The
semi--major axis of the relative orbit of the black hole and companion star,
$\vec{r}_* - \vec{r}_{BH}$, has been calculated with Kepler's law
\begin{equation}
\label{eq2}
4 \pi ^2 a^3_{*-BH} = G P^2 (M_* + M_{BH})\,,
\end{equation}
and then converted to the {\it observable} semi major axis of the orbit of the
companion star projected onto the sky using the relation
\begin{equation}
\label{eq3}
a_* = \frac{1}{d} {M_{BH} \over M_* + M_{BH}}\, a_{*-BH}\,,
\end{equation}
where $d$ denotes the distance of the system from the Sun.
In Eq.~(3) we assume that the accretion disk around the black hole is
optically much fainter than the star.

\begin{table*}[htb]
  \caption{Black hole and companion star parameters}
  \label{tab1}
  \begin{center}
    \leavevmode
        \begin{tabular}[h]{lcccccccc}
	\hline 
	\noalign{\smallskip}
   & $V$ & sp.t. & $M_*$ & $d$ & $P$ & $M_{BH}$ & $i$ & $a_*$ \\
   & mag &  &  $M_\odot$ & kpc & d & $M_\odot$ & deg & $\mu$as \\
	\noalign{\smallskip}
   	\hline	
	\noalign{\smallskip}
Cyg~X-1  & 9 &  O9.7\,Iab & 25  & 2.5 & 5.6  & 13   & 35 & 28 \\
V1003~Sco = GRO\,J1655-40 & 17 & F6\,III  & 2.4 & 3.5 & 2.6 & 6.3  & 70 & 16 \\
V616~Mon = A\,0620-00 & 18 & K4\,V  & 0.7 & 1.2 & 0.32  & 11 & 41 & 16 \\
V404~Cyg = GS\,2023+338 & 19 & K0\,IV & 0.7 & 3.0 & 6.5 & 12 & 56 & 50 \\
V381~Nor = XTE\,J1550-564 & 20 & K3\,III & 1 & 3$^*$ & 1.5 & 10 & 72 & 18 \\
 \noalign{\smallskip}
 \hline	
 \noalign{\smallskip}
 $^*$distance very uncertain \\
       \end{tabular}
  \end{center}
\end{table*}

We conclude from Table~1 that the orbit of the companion star HD\,226868 of
Cyg~X-1 should be observable accurately with Gaia. In addition to the
astrometry, high resolution spectroscopy of the companion star is required in
order to verify the assumption that the disk is faint: The spectral
energy distribution will reveal emission from the disk. Furthermore, light
reflected from the
accretion disk can be detected and measured by Doppler--shifted binary spectral
lines, because the star and the black hole move in opposite directions. Such a
contamination is not
expected for Cyg~X-1. However, in some of the transient X-ray sources light
from the disk could significantly contribute to the system's total light. The
astrometric data must then be reduced as described by Wielen~et~al. (2000).

In the present paper we discuss only those nearby stellar black-hole candidates
that are already known. However, Gould and Salim (2002) point out that there 
is a possibility that Gaia will find many more, if the (presently hypothetical)
``failed supernovae'' -- i.e.~massive stars that directly collapse to black
holes instead of exploding -- do indeed occur.

\section*{Acknowledgements}

We are grateful to Chris Flynn for helpful discussions during the preparation
of this paper.

\end{document}